\documentclass[epsfig]{article}
\usepackage{amssymb}
\usepackage{graphicx}
\usepackage{amsmath}

\input epsf.sty
\input{tcilatex}
\begin{document}

\title{\textbf{F-term Braneworld inflation in Light of Five-year WMAP
observations } }
\author{ \ A.Bouaouda$^{1},$ R.Zarrouki$^{2}$, H.Chakir$^{1,3}$ and M.Bennai$%
^{2,3}$\thanks{%
Corresponding authors: bennai\_idrissi@yahoo.fr, m.bennai@univh2m.ac.ma} \\
$^{1}$Laboratoire de recherche subatomique et applications, {\small Facult%
\'{e} des Sciences \ \ \ \ }\\
\ {\small Ben M'sik, B.P. 7955,} {\small \textit{Universit\'{e} Hassan
II-Mohammedia,}} {\small Casablanca, Maroc. }\\
$^{2}$Laboratoire de Physique de la.Mati\`{e}re Condens\'{e}e{\small , Facult%
\'{e} des Sciences Ben \ \ }\\
\ \ {\small M'sik, B.P. 7955,} {\small \textit{Universit\'{e} Hassan
II-Mohammedia,}} {\small Casablanca, Maroc. }\\
$^{3}${\small Groupement National de Physique des Hautes Energies, Focal
point, LabUFR-PHE, \ }\\
\ \ {\small Rabat, Morocco.}}
\maketitle

\begin{abstract}
We consider a supersymmetric hybrid inflation in the framework of the \emph{%
Randall-Sundrum} type II\textbf{\ }Braneworld model. We drive an analytical
expression for the scalar potential and find that the F-term dominates
hybrid inflation process. We show that for some value of Brane tension we
can eliminate the fine tuning problem related to coupling constant $\kappa $
of the potential$.$ We also calculate all known spectrum inflation
parameters and show that observational bounds from WMAP5, BAO and SN
observations are satisfied.


\end{abstract}

\date{}
\tableofcontents

\bigskip \newpage

\section{Introduction}

Recently there has been considerable interest on supersymmetric hybrid
inflation\cite{SHI} and its Braneworld extension in relation to observation%
\cite{BWHI}. Hybrid inflation was initially introduced to overcome the
shortcomings of chaotic model\cite{linde 90}, but\ cannot explain more
complex questions of inflation theorie, as reheating mechanism due to tachyon%
\cite{SHINF}. It turns out that one may consider hybrid inflation in
supersymmetric theories. In fact, hybrid inflation looks more natural in
supersymmetric theories rather than in non-supersymmetric ones\cite{D.H.Lyth}%
. Thus, it is very interesting to look for a generalized supersymmetric
Braneworld inflation in relation with recent observations\cite{WMAP5}.

In Randall-Sundrum Braneworld senario\cite{L.Randall-R.Sundrum}, our
four-dimensional universe is considered living on a three-dimensional
extended object (brane), embedded in a higher dimensionl space (bulk). We
shall study here some interesting cosmological implications of a
supersymmetric hybrid brane inflation.\bigskip

In this paper, we are interested on F-term effect on perturbation spectrum
in Braneworld inflation. We have considered a Dvali superpotential which
leads to the tree-level potential formed by an F-term and D-term\cite{HBI}.
Note that in most supersymmetric inflationary models only one of this terms
dominates\cite{R.Jocher}. The case of F-term inflation, where F-term
dominates, was considered for the first time in\cite{G.R.Dvali}.

In the present work, we have shown that for some values of Brane tension we
can eliminate the fine tunning problem related to coupling constant $\kappa
. $ We have also analyzed the perturbation spectrum for this potential, in
particular the scalar spectral index, running of spectral index and ratio of
scalar to tensorial amplitude perturbation was calculated. Our results are
in good agreement with recent WMAP5 observations\cite{WMAP5}.

In the next section, we recall first, the foundation of a supersymmetric
version of the hybrid inflation and different perturbation spectrum
expressions in Randall-Sundrum type-II model(RS-II). In the section 3, we
present our result for F-term inflation on the brane. A conclusion and
perspective are given in the last section.

\section{Supersymmetric B\textbf{raneworld Inflation}}

\subsection{Supersymmetric hybrid inflation}

One of the most known one field inflationary models is the power law
potential\cite{linde 90}%
\begin{equation}
V(\sigma )=\frac{1}{2}m^{2}\sigma ^{2}+\frac{1}{4}\lambda \sigma ^{4}
\end{equation}%
where m is associated to scalar field $\sigma .$ This potential is in a good
agreement with observations, in particular for small $\lambda $\cite{liddle
02}.%
\begin{equation}
\left( \frac{\delta T}{T_{0}}\right) _{Q}=6,6\ast 10^{-6}\Leftrightarrow 
\text{ }\lambda =6\ast 10^{-14}\text{ \ et \ }m\lesssim 10^{13}GeV
\end{equation}%
This constraint on the constant $\lambda $, a dimensionless parameter, a
priori in the order of unit, does not seem natural.

The hybrid inflation was introduced to take into account the anisotropie of
temperature for $\lambda $ in the order of unit. The model is based on the
coupling of inflaton $\sigma $ with a second scalar field $\chi $ as\cite%
{Linde94}.%
\begin{equation}
V\left( \sigma ,\chi \right) =\frac{1}{4}\lambda \left( \chi
^{2}-M^{2}\right) ^{2}+\frac{1}{2}m^{2}\sigma ^{2}+\frac{1}{2}\lambda
^{\prime }\sigma ^{2}\chi ^{2}
\end{equation}%
where M is associated to second scalar field $\chi .$ In this model, the
inflaton $\sigma $ is in slow rolling, while the field $\chi $ is
responsible for the destabilization of potential which ends the inflation.
The goal is to be able to take into account the smallness of the CMB
anisotropies with potential coupling constants with more natural values $%
\left( \lambda ,\lambda ^{\prime }\sim 0\left( 1\right) \right) $.
Unfortunately, standard hybrid inflation leads to scalar index spectrum
greater than 1\cite{Linde94}.

In the supersymmetric hybrid Braneworld inflation, it was shown that the
predictions of observable variables match with a recent results\cite{Chafik}%
. In this supersymmetric version, the potential has contributions from
F-term and D-term\cite{F-D-terms}. Note that there exist two classes of
hybrid inflation according to the origin of non-zero contribution of F-term
or D-term to potential.

The F-term inflation is a generalization of the non supersymmetric version
of hybrid inflation described by the potential eq.(3), with only two
parameters. In this model, a superpotential consist of a coupling of two
Higgs superfields $\Phi ,\overline{\Phi }$ and a superfield $S.$ It is given
by\cite{Q. Shafi}%
\begin{equation}
W=\kappa S\left( -\mu ^{2}+\overline{\Phi }\Phi \right)
\end{equation}
$S$ is a scalar superfield singlet under a GUT's \emph{gauge} group which is
the inflaton field. $\kappa $ and $\mu $ are two positive constants. The
superpotential given by eq.(4) is the most general potential consistent with
a continuous R-symmetry under which the fields transform as $S$ $\rightarrow
e^{i\gamma }S$, $W\rightarrow e^{i\gamma }W$ and $\overline{\Phi }\Phi $ is
invariant.

In the following, we consider only the F-term contribution to inflation.
Thus, the scalar potential is derived from the superpotential as%
\begin{equation}
V=\kappa ^{2}\left\vert -\mu ^{2}+\overline{\phi }\phi \right\vert
^{2}+\kappa ^{2}\left\vert S\right\vert ^{2}\left( \left\vert \overline{\phi 
}\right\vert ^{2}+\left\vert \phi \right\vert ^{2}\right) +\text{D-terms}
\end{equation}%
The $\phi $ and $\overline{\phi }$ are \emph{Higgs} fields, which are a part
of the superfields $\Phi $ and $\overline{\Phi }$ respectively$.$ Note that
these fields break U(1) \emph{gauge} symmetry in appropriate
representations. Restricting ourselves to the D-flat direction and bringing $%
S$, $\overline{\phi }$, $\phi $ on the real axis: $S$ $=$ $\sigma $/$\sqrt{2}
$, $\phi $ $=$ $\overline{\phi }$ $=\chi /\sqrt{2}$ where $\sigma $, $\chi $
are normalized real scalar fields, we can obtain the potential which is
similair to hybrid one\cite{A. D. Linde5}. Note that this potential does not
contain the mass-term of scalar fiel $\sigma $ which is of crucial
importance to end inflation.

\bigskip

One way to generate the necessary slope along the inflationary trajectory is
to include the one-loop radiative corrections on this trajectory ($\chi $ $=$
$0$, $\sigma $ $\succ $ $\sigma _{c}=\sqrt{2}\mu $)\cite{Q. Shafi}. In fact,
SUSY breaking by the `vacuum' energy density $\kappa ^{2}\mu ^{4}$ along
this valley causes a mass splitting in the supermultiplets $\phi $, $%
\overline{\phi }$. We obtain a \emph{Dirac} fermion with mass squard equal
to $\frac{\kappa ^{2}\sigma ^{2}}{2}$ and two complex scalars with mass
squard equal to $\frac{\kappa ^{2}\sigma ^{2}}{2}$ $\pm $ $\kappa ^{2}\mu
^{2}$. This leads to the existence of important one-loop radiative
corrections to $V$ on the inflationary valley which can be found from the 
\emph{Coleman-Weinberg} formula\cite{Coleman}$\ $%
\begin{equation}
\Delta V=\frac{1}{64\pi ^{2}}\sum (-1)^{F_{i}}M_{i}^{4}\ln \left( \frac{%
M_{i}^{2}}{\Lambda ^{2}}\right) 
\end{equation}%
where the sum extends over all helicity states $i$, $F_{i}$ and $M_{i}$ are
the fermion number and mass of the i$^{th}$ state respectively, and $\Lambda 
$\ is a renormalization mass scale. The one loop effective potential is then
given by%
\begin{equation}
V_{eff}=\kappa ^{2}\mu ^{4}\left( 1+\frac{\kappa ^{2}}{16\pi ^{2}}\left( \ln
\left( \frac{\kappa ^{2}\sigma ^{2}}{2\Lambda ^{2}}\right) +\frac{3}{2}%
+.....\right) \right) 
\end{equation}%
\ 

Note in passing that the \emph{Coleman-Weinberg }Potential was recently
shown to be in good agreement with WMAP3 observations in the framework of
non-supersymmetric GUTs\cite{shafi}. Here we use this effectif potential to
study perturbation spectrum inflation in Braneworld scenario.\ \ \ \ \ 

\subsection{RS-II Braneworld model}

The Randall-Sundrum Braneworldmodel was mainly studied and shown to give a
very good agreement with recent observations\cite{RS-II OBSERV,ZSB}. In this
cosmological scenario, the metric projected onto the brane is a spatially
fat \emph{Friedmann-Robertson-Walker} with scale factor $a(t)$ and the
Friedmann equation on the brane has the generalized form\cite{P. Binetruy}

\begin{equation}
H^{2}=\frac{8\pi }{3m_{pl}^{2}}\rho \left( 1+\frac{\rho }{2\lambda }\right)
\end{equation}%
where $\rho $ is the energy density of the matter dominted in 3-brane, $%
\lambda $ is the brane tension and $m_{pl}$ is the \emph{Planck} mass.

In Braneworld model, we consider that a scalar field is characterized by an
energy density of the form $\rho =\frac{\dot{\phi}^{2}}{2}+V\left( \phi
\right) .$ The \emph{Klein-Gorden} equation that describes the evolution of
the scalar field is%
\begin{equation}
\ddot{\sigma}+3H\dot{\sigma}+V^{\prime }=0
\end{equation}%
where $\dot{\sigma}=\frac{\partial \sigma }{\partial t}$, $\ddot{\sigma}=%
\frac{\partial ^{2}\sigma }{\partial t^{2}}$ and $V^{\prime }=\frac{dV}{%
d\sigma }.$

We consider the slow-roll approximation ( $\dot{\sigma}^{2}\ll $ $V\left(
\sigma \right) $ and $\ddot{\sigma}\ll $ $H\dot{\sigma}$ ) defined by
parameters $\epsilon $ and $\eta $ given by\cite{Maartens}%
\begin{eqnarray}
\epsilon  &=&\frac{m_{pl}^{2}}{16\pi }\left( \frac{V^{\prime }}{V}\right)
^{2}\frac{\left( 1+\frac{V}{\lambda }\right) }{\left( 1+\frac{V}{2\lambda }%
\right) ^{2}} \\
\eta  &=&\frac{m_{pl}^{2}}{8\pi }\left( \frac{V^{\prime \prime }}{V}\right) 
\frac{1}{\left( 1+\frac{V}{2\lambda }\right) }
\end{eqnarray}%
where $V^{\prime \prime }=\frac{d^{2}V}{d\sigma ^{2}}.$

During inflation $\epsilon \ll 1$ and $\mid \eta \mid \ll 1$. \ The end of
inflation will take place for a field value $\sigma _{end}$ such that $\max
(\epsilon ,\left\vert \eta \right\vert )=1.$ 

The perturbation spectrum of inflation is caracterized by\cite{Maartens}:

The scalar spectral index is presented by%
\begin{equation}
n_{s}\simeq -6\epsilon +2\eta +1
\end{equation}

\bigskip The power spectrum of the curvature perturbations given by%
\begin{equation}
P_{R}\left( k\right) \simeq \frac{128\pi }{3m_{p}^{6}}\frac{V^{3}}{V^{\prime
^{2}}}\left( 1+\frac{V}{2\lambda }\right) ^{3}
\end{equation}

The amplitude of tensor perturbations defined by\bigskip \cite{S. Tsujikawa}%
\begin{equation}
P_{g}\left( k\right) \simeq \frac{128}{3m_{p}^{4}}V\left( 1+\frac{V}{%
2\lambda }\right) F^{2}\left( x\right) 
\end{equation}

\bigskip where $x=Hm_{p}\sqrt{\frac{3}{4\pi \lambda }}$ and $F^{2}\left(
x\right) =\left( \sqrt{1+x^{2}}-x^{2}\ln \left( \frac{1}{x}+\sqrt{1+\frac{1}{%
x^{2}}}\right) \right) ^{-1}.$

\bigskip The ratiooftensorto scalarperturbationsandthe running ofthe
scalarindex presentedrespectively by

\begin{equation}
r\left( k\right) \simeq \left( \frac{m_{p}^{2}}{\pi }\frac{V^{\prime
^{2}}F^{2}\left( x\right) }{V^{2}\left( 1+\frac{V}{2\lambda }\right) ^{2}}%
\right) \mid _{k=k_{\ast }}
\end{equation}%
\begin{equation}
\frac{dn_{s}}{d\ln k}\simeq \frac{m_{p}^{2}}{4\pi }\frac{V^{\prime }}{V}%
\frac{1}{\left( 1+\frac{V}{2\lambda }\right) }\left( 3\frac{\partial
\varepsilon }{\partial \sigma }-\frac{\partial \eta }{\partial \sigma }%
\right) 
\end{equation}

\bigskip Here, $k_{\ast }$ is referred to $k=Ha$, the value when the
Universe scale crosses the \emph{Hubble} horizon during inflation.

Finally, the number of e-folds during inflation is\cite{Maartens}%
\begin{equation}
N\simeq -\frac{8\pi }{m_{pl}^{2}}\int_{\sigma _{\ast }}^{\sigma _{end}}\frac{%
V}{V^{\prime }}\left( 1+\frac{V}{2\lambda }\right) d\sigma 
\end{equation}%
where the subscripts $\ast $ and $end$ are used to denote the epoch when the
cosmological scales exit the horizon and the end of inflation, respectively.

\section{\protect\bigskip \textbf{F-term inflation} in the light of WMAP5}

In this section, we introduce the F-term hybrid inflation by using
Braneworld model. Note that the F-term inflation was introduded to solve the
blue spectrum problem ($n_{s}>1$)\cite{Linde94} and was shown to give a good
values of perturbation spectrum\cite{Chafik}. In the present work we give an
exact evaluation of various inflation parameters. 

The inflation end at $\sigma =\sigma _{c}$ if $\sigma _{c}\geqslant $ $%
\sigma _{end}$. These parameters are given by%
\begin{eqnarray}
\left\vert \eta \right\vert  &=&1\text{ \ \ \ \ \ \ \ \ }\Longrightarrow 
\text{ \ \ \ \ \ \ }\sigma _{end}=\frac{m_{pl}\kappa ^{2}\mu ^{2}}{8\left(
\pi \right) ^{3/2}}\sqrt{\frac{1}{\kappa ^{2}\mu ^{4}\left( \frac{k^{2}\mu
^{4}}{2\lambda }+1\right) }} \\
\frac{\partial ^{2}V}{\partial \chi ^{2}} &\mid &_{\chi =0}=0\text{ \ \ $%
\Longrightarrow $ \ \ \ }\sigma _{c}=\sqrt{2}\mu 
\end{eqnarray}%
this is equivalent to the condition%
\begin{equation}
\sigma _{c}\geqslant \sigma _{end}\Rightarrow \lambda \leqslant \frac{\kappa
^{2}\mu ^{4}}{\frac{m_{pl}^{2}\kappa ^{2}}{64\mu ^{2}\pi ^{3}}-2}
\end{equation}%
thus, we have obtained an upper limit for the brane tension $\lambda $ for
some value of potential parameters $\varkappa $ and $\mu .$

From eq.(17), we evaluate the corresponding inflaton field value $\sigma
_{\ast }$ as%
\begin{equation}
\sigma _{\ast }^{2}\simeq \frac{N\kappa ^{2}m_{pl}^{2}}{32\pi ^{3}\left( 1+%
\frac{\kappa ^{2}\mu ^{4}}{2\lambda }\right) }+2\mu ^{2}
\end{equation}%
On the other hand, the combination of WMAP5, BAO, and SN data gives the
following results\cite{WMAP5}%
\begin{eqnarray}
0.9392 &\prec &n_{s}\prec 0.9986\text{ } \\
r &<&0.20\text{ } \\
-0.0728 &<&\frac{dn_{s}}{dlnk}<0.0087 \\
P_{R}\left( k\right)  &=&2.457\times 10^{-9}
\end{eqnarray}%
From eq.(20) where  $\kappa >\sqrt{128\pi ^{3}}\frac{\mu }{m_{pl}}$, and
taking into account of $\mu =4\times 10^{-4}m_{p}$\cite{Chafik} and $%
m_{pl}=1.2\times 10^{19}GeV$, we derive a numerical limit of a potential
coupling constant $\kappa \gtrsim 0.02.$This result allows us to eliminate
the fine tuning problem.

For $\kappa =0.03$, \ and $N=55$ for example we obtain from eq.(20) an upper
bound for Brane tension as 
\begin{equation}
\lambda \leqslant 5.337\times 10^{-18}m_{pl}^{4}
\end{equation}%
Using $\lambda =2.985\times 10^{-18}m_{pl}^{4}$ which verifies the
inequation (26), we can obtain%
\begin{equation}
P_{R}\left( k\right) \simeq 2.4\times 10^{-9}
\end{equation}%
which is practically the same value provided by the observation.

Using the obtained value of Brane tension $\lambda $, we can estimate the
fundamental \emph{Planck} scale $M_{5}$ 
\begin{equation}
M_{5}\sim 10^{-3}m_{pl}
\end{equation}%
The fundamental \emph{Planck} scale may therefore be significantly below the
effective \emph{Planck} scale. This could provide partially a solution to
the hierarchy problem\cite{N.Arkani-Hamed}.

We note that a large interval of variation of N can reproduce the
observation results as shown in the following table

\begin{center}
\begin{tabular}{|c|c|c|c|}
\hline
\textsl{Inflation parameters} & $n_{s}$ & $r$ & $\frac{dn_{s}}{d\ln k}$ \\ 
\hline
$N=50$ & $0.9610$ & $0.43\ast 10^{-5}$ & $-0.0015$ \\ \hline
$N=55$ & $0.9645$ & $0.39\ast 10^{-5}$ & $-0.0012$ \\ \hline
$N=60$ & $0.9674$ & $0.36\ast 10^{-5}$ & $-0.001$ \\ \hline
\end{tabular}
\end{center}

We can also remark that the scalar spectral index $n_{s}$ and the running of
the scalar spectral index $\frac{dn_{s}}{d\ln k}$ increase according to N,
whereas, the ratio of tensor to scalar perturbations r decreases.

\section{\textbf{Conclusion}}

In this paper,we have studied F-term hybrid inflation in framework of the
Randall-Sundrum Braneworld typeII model. We have derived the effective
potential by introducing the $\emph{Coleman-Weinberg}$ correction to
evaluate various parameters spectrum perturbation. We have shown that the
fine-tuninig problem (very small value of coupling constant $\varkappa $) is
solved and an upper limit for the brane tension $\lambda $ was obtained.
Various spectrum perturbation parameters were also calculated with a
particular choice of $\varkappa $ and $\lambda $. In particular, our results
for power spectrum of the curvature perturbations $P_{R}\left( k\right) $
coincide with recent obsevations. We have also shown a good compatibility of
others inflation perturbation parameters namely the scalar spectral index $%
n_{s}$, the ratio of tensor to scalar perturbations $r$ and the running of
the scalar index $\frac{dn_{s}}{d\ln k}$ with recent observations.

\end{document}